\documentstyle[11pt]{article}
\def \be{\begin{equation}}
\def \ee{\end{equation}}
\normalbaselines
\begin{document}
\bibliographystyle{unsrt}
\begin{flushright}
{TIFR/TH/96-39 \\
July 1996}
\end{flushright}
\vbox{\vspace{5mm}}
\begin{center}

{\LARGE \bf Dilepton asymmetries at $B$ factories in search of
$\Delta B =- \Delta Q$ transitions }\\[7mm] {\bf G. V. Dass}$^1$ and
{\bf K. V. L. Sarma}$^2$\cite{byline}\\[3mm] $^1${\it Department of
Physics, Indian Institute of Technology, Powai, Bombay, 400 076, India
}\\ $^2${\it Tata Institute of Fundamental Research, Homi Bhabha Road,
Bombay, 400 005, India }\\[10mm] \end{center}

\begin{quotation} \small{ 

In order to detect the possible presence of $\Delta B = - \Delta Q$
amplitudes in neutral $B$ meson decays, we consider the measurement of
decay time asymmetries involving like-sign dilepton events at the $B$
factories.

\bigskip

}
  \end{quotation}
  
\baselineskip=0.8cm

Bottom meson decay is a promising place to look for $CP$ violation
outside the neutral kaon system. It will also facilitate a test of the
$CP$ violation mechanism of the standard model \cite{NQ}. One way to
achieve this aim would be to measure decay time asymmetries at the
asymmetric $B$ factories presently under construction \cite{babar}.
But violations of the standard model may show up in $B$ physics in
other ways also. Here we focus on the interesting possibility of the
breakdown of $\Delta B = \Delta Q$ rule in semileptonic decays of
neutral bottom mesons. This rule is not exact as it is violated in
higher order weak interactions. For instance, a $\Delta B = -\Delta Q$
amplitude is generated in the decay $B(\bar bd)\rightarrow \pi
^+e^-\bar \nu_e$ by two tree-level transitions $\bar b \rightarrow
\bar qu \bar d$ and $d\rightarrow qe^- {\bar {\nu }_e}$ by exchanging
the quark $q=u,c,t$. However an estimate of this amplitude may be
misleading because we could have additional contributions (at the same
order and with the same CKM factors) from the $\bar b\rightarrow \bar
d$ Penguin transition, and the effects due to new particle exchanges
in the Penguin loop are unknown
\cite{GOTO}. In any case, an experimental check of the $\Delta B
=\Delta Q$ rule as a phenomenological question is important in its own
right; the rule forms the basis of lepton-tagging of bottom
flavor. For this check, we shall examine the decay time asymmetries
involving like-sign dilepton events produced at asymmetric factories
of $\Upsilon (4S)$.

The parameters of interest are the ratios of semileptonic decay
amplitudes for $B$ and $\bar B$ mesons to decay into the channels
$h_i\ell ^+ \nu_{\ell }$ and the conjugate channels $\bar h_i\ell
^-\bar {\nu}_{\ell }$:
\be 
r_i \equiv {q\over p}{<h_i\ell ^+ \nu_{\ell }|T|\bar B>
\over <h_i\ell ^+ \nu_{\ell }|T|B> }~,~~ 
\bar r_i\equiv {p \over q}{<\bar h_i\ell ^-\bar {\nu}_{\ell }|T|B> 
\over <\bar h_i\ell^-\bar {\nu}_{\ell }|T|\bar B>}~.
\ee
Here $h_i$ stands for a single hadron $h_i= D^-,~D^*(2010)^-~,~\rho
^-~,~...~ $; the constants $p$ and $q$ define the usual
time-propagating states $B_{1,2}=pB\pm q\bar B$ with masses $m_{1,2}$
and widths $\Gamma _{1,2}~.$ We assume complete $CPT $ invariance
throughout, which implies the following relations amongst the
amplitudes
\begin{eqnarray}
 <h_i\ell ^+ \nu_{\ell }|T| B>&=&<\bar h_i\ell ^-\bar {\nu}_{
\ell } |T|\bar B>^*~,\\
\bar r_i&=&\left|{p\over q}\right|^2r_i^*~.
\end{eqnarray}
Note that there is no `strong phase' due to final state interaction as
we are restricting ourselves to semileptonic channels containing a
single hadron (the tiny phase shift due to electroweak scattering is
ignored). The complex parameters $r_{i}$ and $\bar r_{i}$ vanish if
the standard model rule $\Delta B = \Delta Q$ holds.

We concentrate on the exclusive semileptonic decays of the two {\it
neutral} $B$ mesons emitted by $\Upsilon (4S)$ produced at a $B$
factory. Let $\nu (ij)$ denote the number of events obtained by
integrating the decay rate for a beon decaying semileptonically into
the channel $h_i\ell ^+\nu_{ \ell }$ at an instant which will be
identified as $t_i=0$ and the other beon decaying into the channel
$h_j\ell ^+ \nu_{\ell }$ at any subsequent instant $t_j=\tau >0$
\be 
\nu(ij)\equiv \int_0^\infty {\rm Rate}~(h_i \ell ^+\nu _{\ell }~,
~t_i=0~;~h_j\ell ^+ \nu_{\ell }~,~t_j=\tau )~d\tau ~;
\ee 
similarly for the decays into channels $\tilde i$ and
$\tilde j$ we have 
\be 
\nu (\tilde i\tilde j) \equiv  \int_0^\infty {\rm Rate}~(\bar h_i
\ell ^- \bar {\nu }_{\ell }~,~t_{\tilde i}=0~;~\bar h_j\ell ^-
\bar {\nu}_{\ell }~,~t_{\tilde j}=\tau )~d\tau ~.
\ee
Henceforth, for notational brevity we omit listing the leptons and use
the label $i$ of the hadron $h_i$ to identify the {\it channel}
itself, it being understood that the lepton pair ($\ell ^+ \nu _{\ell
}$) accompanies the channel index that has no tilde ($i$) and the
conjugate pair ($\ell ^- \bar {\nu }_{\ell }$) accompanies the channel
index that has a tilde ($\tilde i$); also the decay to the channnel
indexed by the first label always occurs prior to the decay indexed by
the second label.

One can immediately envisage an asymmetry by exploiting the difference
between the numbers $\nu (ij)$ and $\nu (ji)$; in the case of a single
decay channel the difference will have to be between $\nu (i\tilde i)$
and $\nu (\tilde ii)$. Such asymmetries were constructed earlier for
specific lepton charges \cite{DS,SAR92}. In what follows we combine
the production of dilepton events having both $(++)$ and $(--)$ signs;
the resulting asymmetries are more sensitive to the presence of $r_i$
than the previous ones.

We consider the $B^0\bar B^0$ pair (at a $B$ factory) decaying into
exclusive channels $i$ and $j$ and their conjugates $\tilde i$ and
$\tilde j$. From the time-ordering of the two decays, one can measure
$\nu (ij) $ and $\nu (\tilde i\tilde j)$ and get the number of events
having either $(++)$ or $(--)$ dileptons
\be N_{ij}=\nu (ij) + \nu (\tilde i\tilde j)~. \label{enn} 
\ee 
We can therefore form the time asymmetry
\be \alpha (ij)= { N_{ij}-N_{ji} \over N_{ij}+N_{ji}}~.\label{def} 
\ee 
By regarding the parameters $r_i$ , $r_j$ and the $CP$-violation
parameter $(|p|^2-|q|^2)$ to be small, and keeping terms up to first
order of smallness, we obtain
\be \alpha (ij) = -{2y\over 1-a}~{\rm Re}(r_i-r_j)~, \label{asy1} 
\ee 
where the symbols have their usual meaning
\be a= {1-y^2\over 1+x^2}~,~~x={2(m_2-m_1)\over \Gamma
_2+\Gamma _1}~,~~ y={\Gamma _2-\Gamma _1\over \Gamma _2 + \Gamma
_1}~.
\ee 
Using the available experimental value $x= 0.73\pm 0.05$ \cite{PDG}
and assuming that $|y| \ll x$, we see that (ignoring errors)
\be \alpha (ij)
\simeq -5.8~y~{\rm Re}(r_i-r_j)~. 
\ee 
Obviously if experiment shows that $\alpha $ is nonvanishing, at least
one of the parameters $r_i$ or $r_j$ must be nonzero, showing a
breakdown of $\Delta B = \Delta Q$ rule in neutral beon decays. The
case $\alpha =0$ does not lead to a unique conclusion.

The above parameter combination can also be determined by an asymmetry
which involves the oppositely-charged dilepton events
\cite{DS}, but that will be much less sensitive than $\alpha (ij)$
\begin{eqnarray}
{\cal A}_{\ell ^+\ell ^-}(i\tilde j+\tilde ij)&=&-{2y \over 1+a}~
{\rm Re}( r_i-r_j)~, \label{ds}\\ 
&\simeq &0.21 ~\alpha (ij)~.
\end{eqnarray} 
The appearance of the small factor 0.21 is easily understood: while
the like-sign dileptons arise due to $B\bar B$ mixing, the
opposite-sign ones can occur even without mixing and that makes the
total number (denominator) large \cite{fn1}.

A nontrivial variant of $ N_{ij}$ is obtained by choosing the opposite
time-ordering for the $(--)$ dilepton events
\be   N'_{ij}=\nu (ij) + \nu (\tilde j\tilde i)~. \label{ennp} 
\ee
The corresponding asymmetry (obtained by replacing $N$ in
Eq. (\ref{def}) by $N'$) leads to the determination of another
combination of parameters :
\begin{eqnarray} {
\alpha '}(ij)&=&-{2ax\over 1-a}~{\rm Im}(r_i-r_j ) \label{asy2}\\
             &\simeq &-2.7~{\rm Im}(r_i-r_j)~.
\end{eqnarray} 
This asymmetry is interesting as it is not suppressed by the factor
$y$, but requires $CP$ violation in the corresponding decay
amplitudes. It is worth mentioning that the asymmetry $ {\cal A}_{\ell
^+\ell ^-}(i\tilde i)$ associated with unlike-sign dileptons from a
single channel \cite{SAR92}
\be {\cal A}_{\ell ^+\ell ^-}(i\tilde i)= {4ax\over 1+a}~ 
{\rm Im}(r_i)~ \simeq 1.1~{\rm Im}(r_i)~,
\ee 
is far simpler than the above; but its comparison with $\alpha '$ is
not meaningful because the latter involves the parameter-difference
$(r_i-r_j)$.

Lastly, a comment on the effort involved in measuring the above
parameters at a $B$ factory \cite{babar} is in order.  The average
number of events containing same-sign dileptons due to the two
exclusive decay modes ($h_i^{\mp }\ell ^{\pm} \nu $) and ($h_j^{\mp
}\ell ^{\pm} \nu $) is given by
\be 
N_{ij}^T =  \frac{1}{2}~{\cal {L}}~ \sigma ~\chi _d 
~\epsilon _i~\epsilon _j~ f_i~ f_j~ T~,
\ee
Here ${\cal {L}}~(\simeq 10^{34}~ {\rm cm}^{-2}~{\rm s}^{-1})$ is the
luminosity, $\sigma ~(\simeq 1.2~{\rm nb})$ is the $\Upsilon (4S)$
production cross section, factor (1/2) restricts us to neutral beon
pairs, and the $B_d\bar B_d$ mixing ratio $\chi _d~(=0.175\pm 0.016,$
see ref. \cite{PDG}) gives us the event fraction with like-sign
dileptons, and $T$ is the running time.  We shall consider the
interesting case of $h_i=D$ and $h_j = D^{* }(2010)$ for which the
branching fractions are $f_i\simeq 4\%$ and $ f_j\simeq 9\%$
(combining the $e$ and $\mu $ modes) \cite{PDG}; for detection
efficiencies (inclusive of the $D$ and $D^*$ branching fractions to
decay into modes convenient for reconstruction) we shall take as
typical values $\epsilon _i \simeq \epsilon _j \simeq 0.02$. Taking
$T$ to correspond to one full year of working we see that the total
number of likesign dilepton events will be only $ N_d
\sim 48 $.

In general for an asymmetry $A$ the number difference $\delta N$
between two types of events is given by the relation $\delta N=AN$,
where $N$ is the total sample size. We require $\delta N$ to be larger
than the usual Poisson fluctuation $\sqrt{ N}$ (at 1 standard
deviation level). However in practice it may not be easy to meet with
this condition and a small sample of events may only serve to place an
upperlimit on the asymmetry $A<1/\sqrt{N}$, or, a 90\% confidence
level upper limit $A<(1.64/ \sqrt{N}$). Thus with the numbers
mentioned above, one might envisage setting the 90\% CL limits $|y$
Re$(r_i-r_j)| \leq 4\%$ by using Eq. ({\ref{asy1}) and $|{\rm
Im}(r_i-r_j)|\leq 9\%$ by using Eq. ({\ref{asy2}).

In summary, we suggest the measurement of $CP$-conserving asymmetry
Eq.(\ref{asy1}) and $CP$-violating asymmetry Eq.(\ref{asy2}) to test
the validity of the $\Delta B = \Delta Q$ rule in semileptonic decays
of neutral bottom mesons.

\end{document}